\documentclass{IEEEtran}

\usepackage{tikz}
\usetikzlibrary{arrows,shapes}
\usepackage{algorithm}
\usepackage{amsmath}
\usepackage{algorithmic}
\usepackage{subfigure}
\usepackage{graphicx}

\graphicspath{{plots/}}
\begin{document}
\title{Broadcast-based Directional Routing in Vehicular Ad-Hoc Networks}


{
\author{\IEEEauthorblockN{Ahmed Soua, Walid Ben-Ameur, Hossam Afifi}\\
\IEEEauthorblockA{Institut Mines-Telecom, Telecom SudParis, France\\
Email: firstname.name@it-sudparis.eu}}
}

\maketitle

\begin{abstract}
An optimized broadcast protocol is proposed for VANETs. It is based
on two key information: the direction to the destination and
the beamforming angle $\theta$.\\
The efficiency of this technique is demonstrated in terms of packet
delivery, bandwidth gain and probability of transmission success.\\
An analytical model is developed to calculate the transmission area.
This model allows capturing the propagation shape of the forwarding
area. Comparisons with simulations show that the analytical model is
precise.\\

\end{abstract}

\begin{keywords}
Broadcast, Vehicular Ad Hoc Networks (VANETs), angle of
transmission, area of transmission, bandwidth, probability of
success, number of hops.
\end{keywords}

\IEEEpeerreviewmaketitle

\section{Introduction}

Vehicular ad hoc networks have been essentially developed  to
enhance driver safety. We believe that to efficiently implement this
kind of systems a major problem needs to be solved: "how to rapidly
disseminate the information among the vehicles?". At the wireless
link level, a step has been already made in the 802.11 standard [15]
by accelerating the link setup (the IEEE 802.11p removes the need of
association between stations before communicating). At the network
level, one method that is considered to be very effective is
broadcast. It is a technique that does not require a prior
end-to-end connection establishment or maintenance. It is cheap and
simple in terms of deployment
and gives good performance thus its popularity.\\

In this paper, we propose an Optimized Broadcast-based Directional
Routing (OBDR) method that combines broadcast and position-based
routing. OBDR disseminates the information on a set of chosen nodes
based on their distance from their selector and the angle they have
with it and with the destination. Our broadcast mechanism can
replace routing and hence be the only system needed to send
information, or it can be implemented at the physical layer with
directive antennas. It is, in this case, complementary to routing in
upper layers. It can be combined for example with other solutions
such as the one of [12] where a geographical routing mechanism is
proposed.

Another major contribution in our paper is the development of an
analytical model providing a very close approximation of the total
transmission area in the broadcasting scheme. Simulation results
confirm the quality of the model.

\section{Related work}\label{stateArt:sec}

In vehicular communications we mainly find two major approaches for
routing protocols: topology-based and position-based routing
protocols. In topology-based routing, each node has to keep track of
the exact topology of the network in order to compute exact routes
to potential destinations. DSR [1] and AODV [2] are the source of
several topology-based protocols that have been proposed. These
protocols were designed for MANETs and then adapted to VANETs but
they still suffer from many limitations especially in highly dynamic
networks.

In order to remediate to the deficiencies of topology-based
protocols, position-based routing protocols [3] were introduced.
They assume that each vehicle is aware of the positions of other
vehicles in order to select its one hop neighbor that ensures the
communication between the nodes even in high mobility scenarios.
This type of routing has several advantages namely the low overhead
since no prior establishment and maintenance of routes is required
and the good performance in highly dynamic scenarios.

Many position-based protocols have been proposed. GPSR is one
example that consists of forwarding a packet to the one-hop neighbor
that minimizes the distance to the destination [4]. Y. Wang and al.
in [5] propose an improvement of the GPSR protocol by enhancing the
decision-making of data delivery. In fact, they use the concept of
vector, called also Greedy mode, to choose the next relay to enhance
the accuracy of the routing scheme. For intersections scenarios,
they add a predictive mode to predict the motions of neighboring
vehicles. Simulations reveal that their technique outperforms GPSR
protocol in terms of packet delivery ratio and routing overhead.
In [6], authors highlight the importance of the direction of the
moving nodes on the lifetime of the links between a node and its
neighbors.

We note that the direction of the destination hasn't been taken into
consideration in most of the previous algorithms. In [7] authors
propose a Position-based Directional Vehicular Routing algorithm.
Next hops are selected based on their angular directions relative to
the destination. Note that all of the above routing algorithms try
to ensure point-to-point communications and does not focus on
broadcast.

In [8], a role-based multicast protocol is proposed. It suppresses
broadcast redundancy by assigning shorter waiting time prior to
rebroadcasting to more distant receivers. This study focuses on
achieving maximum reachability in a sparsely connected or fragmented
network. To deal with the effect of node density and wireless
channel quality, Slavik and al. propose  a multi-hop wireless
broadcast scheme for vehicular networks called DADCQ [9]. This
protocol utilizes the distance method to select forwarding nodes. In
addition, they adapt their decision threshold function to the
density of nodes, the node clustering factor and the Rician fading
parameter. DADCQ achieves high reachability and low bandwidth
consumption compared to other existing multi-hop dissemination
protocols.

Sun, Feng and al. in [17] proposed the TRAck Detection protocol
(TRADE) which classifies the neighboring vehicles into three main
categories according to their position on the road; same road-ahead,
same road behind and different road. Then, the algorithm selects
nodes from each group in order to forward safety messages; the
farthest vehicles from same road-ahead and same road-behind and all
vehicles from different road are chosen to rebroadcast the alert
message.

Claudio E. Palazzi and al.[10] also proposed the Fast Broadcast (FB)
protocol which uses a distance-based approach with an estimated
transmission range in order to reduce the number of redundant
transmissions of the alert message as well as the hops to be
traversed. This scheme is composed of two phases. The first one,
named estimation phase, aims to provide each vehicle with an
up-to-date estimation of its backward transmission range. On the
other hand, the second one, called broadcast phase, is performed
only when a message has to be broadcasted to all cars in the
sender's area-of-interest.

In [11], authors propose a zone forwarding scheme for information
dissemination in vanets (ZBF). The proposal tackles the problem of
retaining the message in the target zone for the duration of the
effect time. Thus, the algorithm divides the effect area of the
alert message into segments of length the transmission range. A
forwarder, a vehicle which is elected in each segment, is assigned
the task of broadcasting the information to other neighboring
vehicles. When the forwarder is about to leave the zone, a new
forwarder will be elected. Simulation results show that ZBF
outperforms other information dissemination protocols.

To deal with the broadcast issue at road intersections, Tung et al.
propose a directional broadcast protocol, by using precise GPS
navigation system, called Efficient Road-based Directional broadcast
(ERD) [13]. It groups vehicles based on their road segments (single
road case, node near an intersection case and node in an
intersection) and selects relay nodes for each group. The protocol
improves packet efficiency significantly but it still depends on a
very precise position navigation system and an accurate road digital
map which are not always present on each vehicle.

Finally, in [14], the authors present an area-based broadcast
technique called Location-Based Directional Broadcast protocol
(LDB). It relays the alert message taking into account the position
and velocity of the concerned vehicles. The authors give some
heuristic rules to choose the best relay nodes.

Despite the acceptable performances, we believe that these
algorithms are too complex for the context and the purpose of
vehicular safety applications. Moreover, they rely on precise
navigation positioning system which is not available on all
vehicles.

None of the previous proposals try to derive analytical models for
their system. In our study, we design a novel solution to reduce the
bandwidth usage by limiting the numbers of broadcasters. Our
proposal, compared to previous state of the art, is much simpler and
does not require precise positioning system. Rather than using
simulation, we derive analytical models for our technique to
estimate performance metrics such as probability of success,
bandwidth gain and implicated nodes. We believe these new models
will be helpful to analyze similar problems and that on the long
run, they will replace simulation tools.

\section{PROPOSED TECHNIQUE}

We suppose that a message is sent from a source to a destination in
a vehicular context. The nodes (vehicles) are assumed to be randomly
positioned in an area. The nodes that receive an intermediate
message will broadcast it on their turn unless they have seen it
before, or unless they are the destination. There is no knowledge on
whether the message has reached its destination or not. This means
that a node may continue to broadcast while the message has already
arrived. The system ends broadcasting when there are no more relays
to broadcast. In this case, we assume that the message has arrived
or that the system has encountered a transmission failure. The
transmission failure is a situation when there is not any
intermediate relay to reach the destination.

The main feature of our protocol design is that the broadcast
operation will be adjusted based on some basic geographic
information and that it will use a directive beamforming system.
\\
The simple geographic information is composed of two parameters.

\begin{figure}[!htbp]
  \centering
  \includegraphics[height=4.5cm,width=8.44cm]{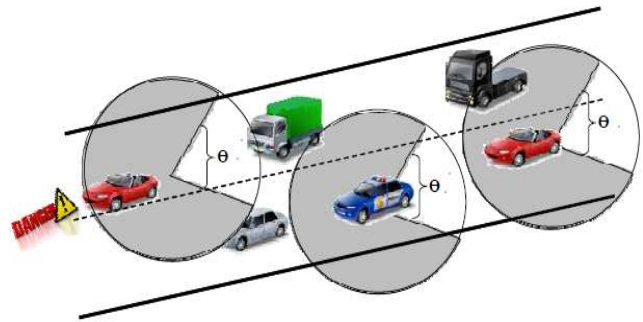}
 \caption{Directional beamforming broadcast scheme}

\end{figure}

The first is  the destination direction (it can be an approximate
value). Our algorithm can be adapted to relatively precise or less
precise direction information as it will be seen. The success or
failure of the transmission is strongly
dependent on the density of neighbor nodes $\lambda$ .\\
We believe that the general information of direction in the
vehicular context is very meaningful. In general, we can see that
communication is either directed to the roadside or to cars coming
or going in the same or in the opposite directions.

The second parameter is the beamforming angle $\theta$ (see Figure
1). This angle has a great effect on the overall performance. A very
wide angle provides some insurance that the message would reach its
destination. Notice that when the information related to the
direction of the destination is uncertain, using a large angle
provides some robustness. However, a larger set of nodes is involved
and the performance of the overall relaying system might be bad. \\
A small beamforming angle prevents waste of bandwidth
but may lead to more transmission failures.

\section{Results}

In this section, we evaluate the performance of our proposed
technique through MATLAB simulations. We study precisely the impact
of the density of nodes and the beamforming angle on the
performances of our proposal. The results confirm the efficiency of
our algorithm in terms of use of bandwidth, the ratio of implicated
nodes and the probability of transmission success.

In fact, transmitting messages within a limited area clearly reduces
the number of nodes having to relay the message and considerably
reduces the total number of transmissions and affect the probability
of success. Moreover, varying the angle of transmission $\theta$ and
the density of the nodes will give us an indication on the spectrum
efficiency.

There will be two kinds of results in this section. Some simulation
results related to the probability of success and the proportion of
vehicles participating to the forwarding process. An analytical
model will also be given for the second performance criteria (the
proportion of implicated vehicles).

\subsection{Some simulation results}

In order to analyze the performance of our protocol, we conduct
several simulations varying the following system parameters:

\begin{itemize}
\item Number of vehicles in the simulation from 1000 to 3000.
\item Beamforming angle $\theta$ (from $22.5^{\circ}$ to $135^{\circ}$).
\end{itemize}

The nodes are assumed to be randomly positioned in a square area
according to a Poisson process of density $\lambda$ (so $\lambda$ is
equal to the number of nodes divided by the area of the square).\\
This assumption can safely be adopted since during rush hour traffic
in an urban scenario the velocity of vehicles is
limited and the vehicle density is very important.\\
Indeed, the speed of vehicles in an urban area, compared to the
duration of wireless transmission, is very small and therefore we
can assume that the node positions and connectivity
do not change significantly from a broadcast operation to broadcast operation.\\
Furthermore, in such urban conditions (large density of vehicles,
small inter-vehicle spacing) terrain is not a significant factor.

In Figure 2, we show a snapshot of the nodes implicated in the
transmission of a message from source \emph{S} to destination
\emph{D}. We consider that the angle is
    $\theta=60^{\circ}$
and that all nodes have a transmission range 200m. The figure shows
that the relays are located within a geographical area.  The
analytical model will prove that this shape is similar to a leaf.
\begin{figure}[!htbp]
  \begin{center}
  \includegraphics[height=5cm,width=7.44cm]{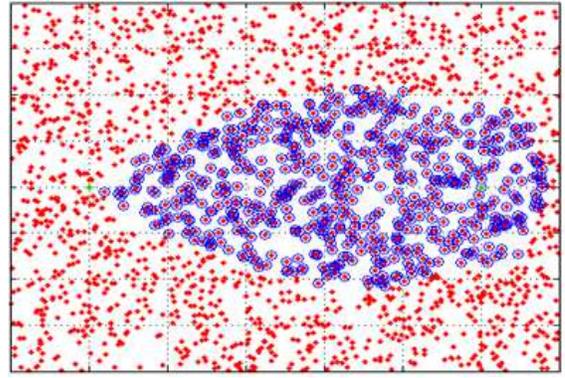}
  \end{center}
  \caption{Network topology and distribution of implicated nodes in the transmissions}
\end{figure}

Figure 3 gives the ratio of implicated nodes in the transmission
when varying the angle $\theta$. Clearly, the fact that we transmit
messages within a limited area reduces the number of nodes having to
relay the message and in turn
considerably reduces the total number of transmissions.\\
In addition, the use of indicators to determine whether a node had
already transmitted a message gives us the possibility to avoid
unnecessary retransmissions and hence avoid the broadcast storms.

\begin{figure}[!htbp]
  \begin{center}
  \includegraphics[height=6cm,width=8.54cm]{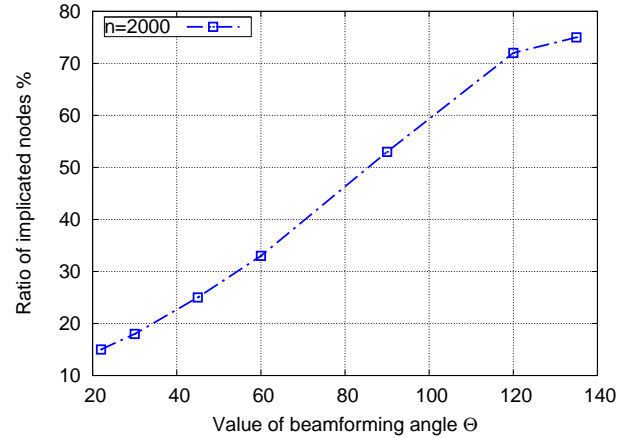}
  \end{center}
  \caption{Ratio of implicated nodes}
\end{figure}
Figure 4 shows the gain in bandwidth (and hence the spectrum
efficiency) when varying the angle of transmission $\theta$ with a
total number of nodes equal to 2000. The spectrum use ratio of
relays is presented by the product of the implicated nodes' ratio
and
    $\frac{\theta}{360^{\circ}}$
.

\begin{figure}[!htbp]
  \begin{center}
  \includegraphics[height=6cm,width=8.54cm]{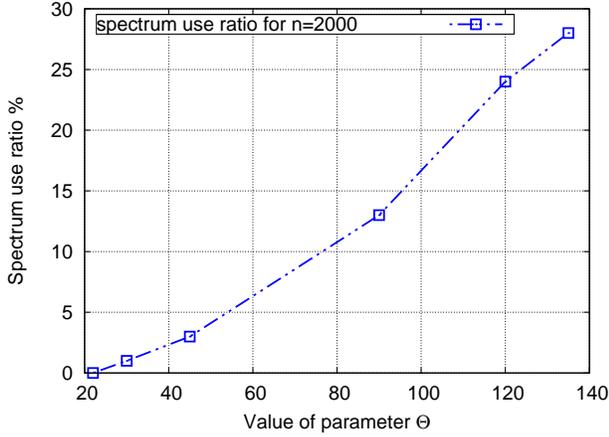}
  \end{center}
  \caption{Gain in bandwidth}
\end{figure}

\begin{figure}[!htbp]
  \begin{center}
  \includegraphics[height=6cm,width=8.54cm]{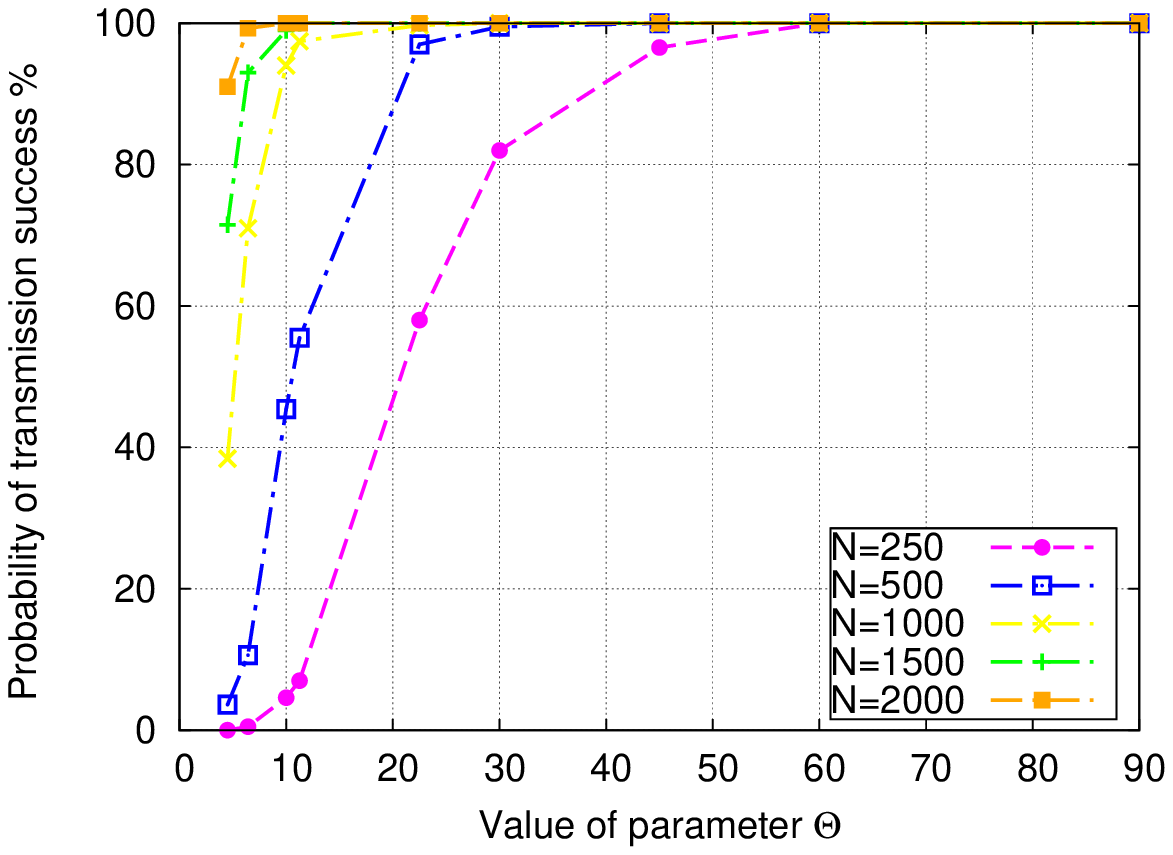}
  \end{center}
  \caption{Probability of transmission success (d=1000m)}
\end{figure}

\begin{figure}[!htbp]
  \begin{center}
  \includegraphics[height=6cm,width=8.54cm]{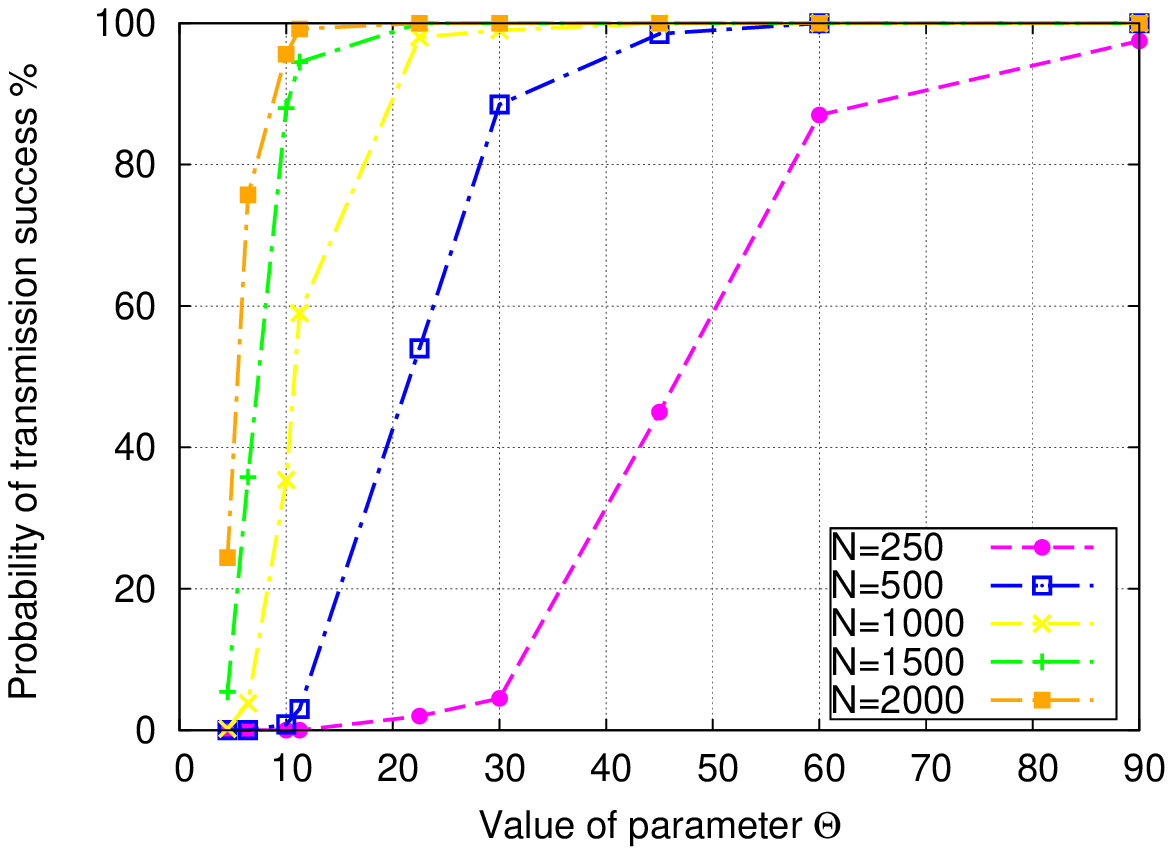}
  \end{center}
  \caption{Probability of transmission success (d=2000m)}
\end{figure}

\begin{figure}[!htbp]
  \begin{center}
  \includegraphics[height=6cm,width=8.54cm]{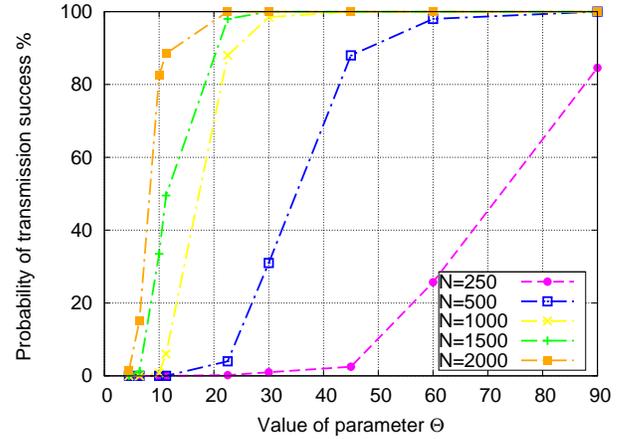}
  \end{center}
  \caption{Probability of transmission success (d=3000m)}
\end{figure}

Observe that the beamforming mechanism achieves a large gain in the
spectrum use when compared to a typical broadcast operation
involving all nodes.

Figures 5, 6 and 7 show the variations of the success probability
with the beamforming angle $\theta$.\\
In each figure, different densities are considered (different node
numbers). The distance \emph{d} between the source and the
destination is equal to 1000m in Figure 5, while it equals 2000m
(resp.3000m) in Figure 6 (resp.7).\\
Observe that the probability of success seems to be convex for small
angles and concave for large angles. \\
One can deduce from Figures 5, 6 and 7 that the angle should be
chosen carefully depending on the distance and the density.

\subsection{An analytical model for the transmission area}

In this section we present an analytical model to estimate the
transmission area (.i.e., the set of nodes that receive a message
and retransmit it). \\
We assume that nodes are uniformly distributed in a square. The
source and all relay nodes transmit messages according to an angle
$\theta$ within a radius \emph{r}.

As shown on Figure 8, the transmission can be approximated by a set
of triangles (the destination is a vertex of each triangle).
Starting from the triangle containing the source, each two
consecutive triangles have a common vertex (in addition to the
destination). Each triangle's vertex is obtained by considering that
there is a mobile situated in a vertex of the previous triangle
transmitting within a distance \emph{r} according to an angle
$\theta$. It should be understood that the union of these triangles
is only an approximation of the true transmission area. We will see
that this approximation is very good.  Observe that the transmission
area is again similar to a leaf.

\begin{figure}[!htbp]
  \begin{center}
  \includegraphics[height=4.5cm,width=8.44cm]{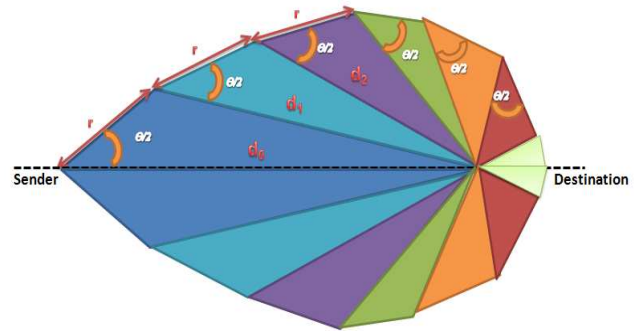}
  \end{center}
  \caption{Area of transmission}
\end{figure}

Since the transmission area is symmetric, we can focus on the half
plane above the line connecting the source and the destination.\\
Starting from the triangle containing the source and using $d_{i}$
to denote the length of the common edge between the triangle number
\emph{i} and the triangle number
    $i+1$
, the area of triangle number \emph{i} is given by:
\begin{equation}\label{eq1}
    S_{i}=\frac{1}{2} \emph{r} d_{i} \sin{\frac{\theta}{2}}
\end{equation}

Moreover, the distances $d_{i}$ and $d_{i-1}$ are linked through the
following equation:
\begin{equation}\label{eq2}
    d^{2}_{i}=d_{i-1}^{2}+r^2-2r d_{i-1}\cos{\frac{\theta}{2}}
\end{equation}

All these distances and triangle areas can be iteratively computed
using the two previous equations. This directly leads to the total
area given by:
\begin{equation}\label{Eq.3}
    S=2\sum _{i=1}^{N_{b}}S_{i}
\end{equation}

We evaluate hereafter the model in \eqref{Eq.3} and compare it to
our simulations. The result is expressed in terms of the ratio of
implicated nodes with respect to the angle $\theta$.

We show on Figure 9 the ratio of implicated nodes given by
simulation when there are 1000 nodes in the square and also when
there are 3000 nodes. We also show the theoretical ratio given by
Eq.(3). Notice that the theoretical value does not depend on the
mobiles density.
\\
One can see that the three curves are very close. This clearly
implies that the analytic model provides a very good approximation
of the real situation.

\begin{figure}[!htbp]
  \begin{center}
  \includegraphics[height=6cm,width=8.54cm]{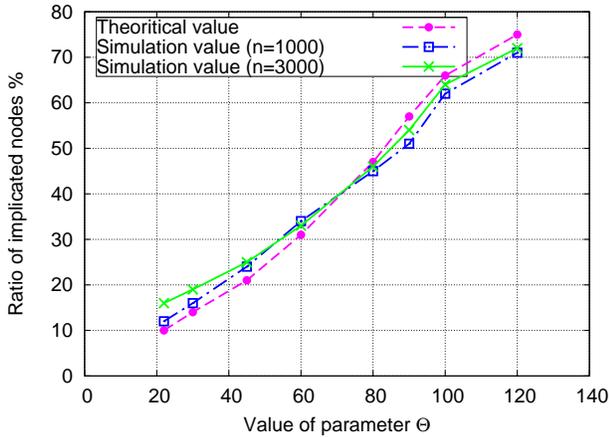}
  \end{center}
  \caption{Comparison between the theoretical and practical values of the implicated nodes ratio for different densities}
\end{figure}

\begin{figure}[!htbp]
  \begin{center}
  \includegraphics[height=6cm,width=8.54cm]{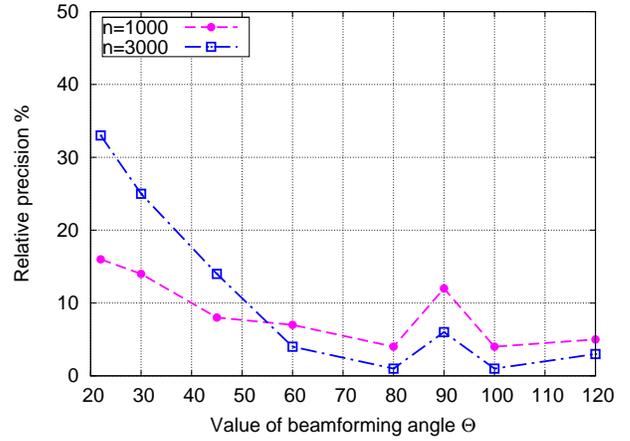}
  \end{center}
  \caption{Relative simulation results}
\end{figure}

Figure 10 represents the relative error of the model computed as
follows:

\begin{equation}\label{Eq.4}
    \emph{relative}_{error}=\frac {theoratical_{value}-
    simulation_{value}}{simulation_{value}}
\end{equation}
The relative error seems to be particularly small for angles between
$45^{\circ}$ and $120^{\circ}$.\\
An important observation related to Figure 9 (and confirmed by many
other simulations) is that the ratio of implicated nodes seems to
vary linearly when $\theta$ is less than $100^{\circ}$.

\section{conclusion}
In this paper, we proposed a technique combining geographic routing
and broadcast. The nodes located at a distance less then \emph{r}
from the emitting vehicle and belonging to the angular sector of
opening angle $\theta$ in the direction of the destination are
chosen as relays.

As mentioned before, OBDR can serve as a substrate to overlay
systems. In fact, it can be applied as a prior phase of a routing
technique such as ant and swarm approaches.

An analytical model was given to estimate the ratio of vehicles that
will relay the message. Comparisons with simulations have clearly
shown the very good precision of the model. More simulations have
been conducted to estimate the probability of a message to reach the
destination. This probability is high for typical values of vehicle
density and distances between the source and the destination.

This combination of beamforming technique and geographic routing
suggests that OBDR should be able to send alert messages with some
guarantees in terms of success probability and without overloading
the network and wasting bandwidth for vehicular urban environment.

\end{document}